\documentclass[twocolumn, pre, showpacs, english, preprintnumbers, amsmath, amssymb, superscriptaddress, aps,longbibliography]{revtex4-2}
\usepackage[subpreambles=true]{standalone}
\usepackage{comment}
\usepackage[utf8]{inputenc}
\usepackage{graphicx}
\usepackage{grffile}
\usepackage[usenames,dvipsnames]{xcolor}
\usepackage{amsmath}
\usepackage[normalem]{ulem}
\usepackage[resetlabels, labeled]{multibib}
\usepackage{appendix} 

\newcites{S}{References Supplementary Materials}
\definecolor{orange}{rgb}{1,0.5,0}
\definecolor{goodgreen}{rgb}{0.1,0.5,0}
\definecolor{goodred}{rgb}{0.7,0,0}
\usepackage{lineno}
\renewcommand\vec{\boldsymbol}

\usepackage{tikz}
\usepackage{tocvsec2}
\usepackage{scrextend}
\makeatletter

\usepackage[colorlinks,urlcolor=goodgreen,citecolor=blue,linkcolor=goodred]{hyperref}

\usepackage{color}

\usepackage{float}
\graphicspath{ {figures/} }

\let\oldepsilon\epsilon \let\epsilon\varepsilon \let\varepsilon\oldepsilon

\makeatother

\usepackage{babel}

\begin{document}

\title{Positive magnetoresistance in anapole superconductor junctions}
\newcommand{\orcid}[1]{\href{https://orcid.org/#1}{\includegraphics[width=8pt]{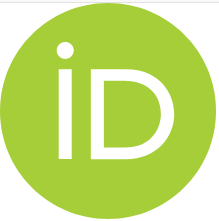}}}

\author{Tim Kokkeler\orcid{0000-0001-8681-3376}}
\email{tim.kokkeler@dipc.org}
\affiliation{Donostia International Physics Center (DIPC), 20018 Donostia--San Sebasti\'an, Spain}
\affiliation{University of Twente, 7522 NB Enschede, The Netherlands}

\author{Alexander Golubov\orcid{0000-0001-5085-5195}}
\affiliation{University of Twente, 7522 NB Enschede, The Netherlands}
%\author{Sebastian Bergeret\orcid{0000-0001-6007-4878}}
%\affiliation{Centro de F\'isica de Materiales (CFM-MPC) Centro Mixto CSIC-UPV/EHU, E-20018 Donostia-San Sebasti\'an,  Spain}
%\affiliation{Donostia International Physics Center (DIPC), 20018 Donostia--San Sebasti\'an, Spain}
\author{Yukio Tanaka\orcid{0000-0003-1537-4788}}
\affiliation{Deparment of Applied Physics, Nagoya University, 464-8603 Nagoya, Japan\\
}
\affiliation{Research Center for Crystalline Materials Engineering, Nagoya University, 464-8603 Nagoya Japan}
%\begin{document}
\begin{abstract}
    The article presents a method to detect time-reversal symmetry breaking in non-centrosymmetric superconductors using only transport measurements. Specifically, if time-reversal symmetry is broken via a phase difference between singlet and triplet correlations, as in anapole superconductors, the conductance in SFN junctions is enhanced by increasing the exchange field strength in the ferromagnet. 
    This is in sharp contrast with the negative magnetoresistance when using superconductors in which time reversal symmetry is preserved.
    Moreover, results show a large quadrupolar component of the magnetoresistance which is qualitatively different from the bipolar giant magnetoresistance in strong ferromagnets.
    %Moreover, we show that the dependence of this effect on the angle of the exchange field is different than for ferromagnets. With this we provide an extra tool to identify the pair potential in unconventional superconductors.
\end{abstract}
\maketitle
%\onecolumngrid\
\section{Introduction}
Recent advances within superconductivity focus on the understanding of unconventional superconductors \cite{sigrist1991phenomenological,schnyder2008classification,kallin2009sr2ruo4,maeno2011evaluation,sigrist2005review}, i.e. superconductors which do not obey BCS theory. 
In unconventional superconductors triplet \cite{mackenzie2017even,kallin2016chiral,linder2019odd,balian1963superconductivity,sigrist1991phenomenological,fu2008superconducting,chiu2021observation,bauer2004heavy,linder2015superconducting,suh2020stabilizing,aoki2019review, saxena2000superconductivity,aoki2001coexistence,hardy2005p,huy2007super,ran2019nearly,yang2021spin,tanaka2004anomalous,tanaka2005theory} and odd-frequency \cite{berezinskii1974new,balatsky1992new,belitz1992even,abrahams1995properties,kirkpatrick1991disorder,coleman1997reflections,linder2019odd,cayao2020odd,gentile2011odd,tanaka2007theory,tanaka2011symmetry,tsintzis2019odd,kuzmanovski2020suppression} pairings may appear.
Most attention is paid to superconductors in which inversion ($\mathcal{P}$) and time-reversal ($\mathcal{T}$) symmetry are preserved in the pair potential. However, several known superconductors  break time-reversal symmetry, either in bulk or near the edges 
\cite{wenger1993dwave,rokhsar1993pairing,covington1997observation,sigrist1991phenomenological,sigrist1991phenomenology,laughlin1994tunneling,laughlin1998magnetic,hillier2009evidence,wakatsuki2017nonreciprocal,kivelson2020proposal,suh2020stabilizing,xia2006high,luke1998time,clepkens2021shadowed,sigrist1998time,ghosh2020recent,ghosh2022time,lee2009pairing,farhang2023revealing,ajeesh2023fate,shang2020simultaneous,grinenko2021split,willa2021inhomogeneous,maisuradze2010evidence,roising2022heat,movshovich1998low,balatsky1998spontaneous,belyavsky2012chiral,black2014chiral,tanaka2001influence,sigrist1995fractional,palumbo1990magnetic,kuboki1996proximity,tanuma1999quasiparticle,tanuma2001tunneling,fogelstrom1997tunneling,krupke1999anisotropy,matsumoto1995coexistence,kallin2016chiral}, 
while in superconductors in which the crystal structure breaks inversion symmetry, the pair potential may contain both even and odd-parity components 
\cite{bauer2004heavy,amano2004superconductivity,akazawa2004pressure,togano2004Superconductivity,tateiwa2005novel,kimura2005pressure,sugitani2006pressure,honda2010pressure,SETTAI2007844,Bauer2010Unconventional, xie2020captas,bauer2012non}. Also the breaking of inversion symmetry near an edge may cause a local admixture of even-parity and odd-parity superconductivity \cite{tanaka2004theory,suzuki2022destruction}.  

Moreover, in a recently discussed class of superconductors, anapole superconductors, both symmetries are broken \cite{kanasugi2022anapole,kitamura2022quantum,chazono2022piezoelectric,goswami2014axionic,mockli2019magnetic}. Examples of possible anapole superconductors are $\text{UTe}_{2}$ \cite{kanasugi2022anapole,kitamura2022quantum,chazono2022piezoelectric}, $\text{Cu}_{x}\text{Bi}_{2}\text{Se}_{3}$ and $\text{Sn}_{1-x}\text{In}_{x}\text{Te}$ \cite{goswami2014axionic}, while also non-centrosymmetric superconductors with a magnetic ordering in their phase diagram, such as $\text{CePt}_{3}\text{Si}$ \cite{bauer2004heavy},   UIr \cite{akazawa2004pressure}, $\text{CeRh}_{2}\text{As}_{2}$ \cite{mishra2022anisotropic} and $\text{CeCu}_{2}\text{Si}_{2}$ \cite{wu2021revealing},  are promising platforms for $\mathcal{P}$ and $\mathcal{T}$ broken superconducting phases in some parameter regimes. Next to this, time-reversal symmetry can be broken via the inverse proximity effect of a ferromagnet or ferromagnetic insulator \cite{bergeret2004induced}. Thus, both non-centrosymmetric and time-reversal symmetry broken superconductivity is locally abundant.
The proximity effect of superconductors that are time-reversal and/or inversion symmetry broken can be significantly different from the proximity effect of superconductors in which these symmetries are preserved \cite{iniotakis2007andreev,eschrig2010theoretical,annunziata2012proximity,rahnavard2014magnetic, mishra2021effects,ikegaya2021proposal,daido2017majorana,chiu2023tuning,ishihara2021tuning,kokkeler2023proximity}.

Superconductors in which both time reversal symmetry and inversion symmetry are broken have great potential for future applications, for example in non-reciprocal transport, since the conditions for non-reciprocal transport to occur, time-reversal symmetry and gyrotropy \cite{silaev2017anomalous,zhang2022general}, are met intrinsically in the bulk material. Thus, such superconductors are a promising platform for superconducting diodes \cite{wakatsuki2017nonreciprocal,wakatsuki2018nonreciprocal,ando2020observation,pal2022josephson,baumgartner2022supercurrent,hou2022ubiquitous,lin2022zero,lyu2021superconducting,wu2021realization,silaev2014diode,yuan2022supercurrent,tanaka2022theory,pal2019quantized,mayer2020gate,bauriedl2022supercurrent,
kopasov2021geometry,margaris2010zero,chen2018asymmetric,he2022phenomenological,kokkeler2022fieldfree,karabassov2023superconducting,de2023superconducting,banerjee2023enhanced,cuozzo2023microwave,vigliotti2023reconstruction,ilic2022theory,ilic2022current,souto2022josephson,he2023supercurrent,maiani2023nonsinusoidal,de2023superconducting,zazunov2023approaching,nadeem2023superconducting,lu2023superconductivity,lu2023tunable,costa2023microscopic,davydova2022universal,karabassov2022hybrid,dolcini2015topological,daido2022intrinsic,ciaccia2023gate,song2023interference,nadeem2023superconducting,kokkeler2023nonreciprocal,tanaka2022giant}, no external source of an exchange field and/or spin-orbit coupling is needed. This greatly simplifies the setup for such diodes.

The proximity effect and transport properties of non-centrosymmetric or time reversal symmetry broken superconductors has been studied in detail in several limits \cite{iniotakis2007andreev,eschrig2010theoretical,annunziata2012proximity,rahnavard2014magnetic, mishra2021effects,ikegaya2021proposal,daido2017majorana,chiu2023tuning,asano2011josephson,klam2014josephson,wu2009tunneling,fujimoto2009unamiguous,borkje2006tunneling,tanaka2010anomalous,yada2011surface,sato2011topology}. 
Recently, a dirty limit transport theory was discussed, focussing on (i)s + p-wave superconductors \cite{tanaka2022theory,kokkeler2023proximity}. In such superconductors the pair potential breaks inversion symmetry as indicated by the simultaneous presence of even-parity s-wave and odd-parity p-wave components, while time-reversal symmetry is broken if there is a finite phase difference between the singlet and triplet correlations.  %It has been shown that the conductance in SNN junctions with (i)s + p-wave superconductors strongly depends on the phase difference between the s and p-wave components of the pair potential \cite{kokkeler2023proximity}. 
In \cite{kokkeler2023proximity} the importance of considering time-reversal symmetry breaking of the pair potential of superconductors was illustrated using SNN junctions. 
However, in SNN junctions the dependence of the conductance on the phase between the singlet and triplet components can not be unambiguously distinguished from the dependence on the ratio of their magnitudes. Therefore, a method to obtain smoking gun evidence for time-reversal symmetry breaking based only on transport measurements is so far absent.
%However, in SNN junctions the effects of a change in the phase difference between the singlet and triplet pair amplitudes can not be well distinguished from the effect of a change in the ratio of their magnitudes. A smoking gun evidence for time-reversal symmetry breaking based only on transport measurements is so far absent.
%However, the conductance is similarly influenced by the ratio between the s-wave and p-wave components, and therefore a smoking gun ev

In this paper we provide such method by showing that time-reversal symmetry breaking in the superconductor can be identified using SFN junctions. We show that for the time-reversal symmetry broken non-centrosymmetric is + p-wave superconductors the conductance for voltages just below the superconducting gap can be strongly enhanced by an exchange field. This is in contrast with the negative magnetoresistance in non-centrosymmetric superconductors that obey time-reversal symmetry such as s + p-wave superconductors. 

%The positive magnetoresistance is maximized if the exchange field is perpendicular to the d-vector of the p-wave correlations. 
Moreover, the dependence of the conductance enhancement on the exchange field direction provides an additional tool to determine the direction of the d-vector of the p-wave correlations. We show that if the exchange field strength $h$ is much smaller than the Fermi energy $E_{F}$ ($h/E_{F}\ll 1$), the magnetoresistance is quadrupolar, with maxima when d-vector and exchange field are perpendicular. Thus, for ferromagnets with $h/E_{F}\ll 1$ the anisotropy of the magnetoresistance is qualitatively different from the well-known bipolar magnetoresistance for $h/E_{F}\sim 1$ \cite{gijs1997perpendicular,bozovic2002coheren,miyazaki1995giant,binasch1989enhanced,gijs1993perpendicular,pratt1991perpendicular,bauer1992perpendicular,valet1993theory}. %\TK{I STILL NEED TO ADD CITATIONS HERE.}

Next to this, we show that the voltage window in which the conductance is enhanced is determined by the ratio of the s-wave and p-wave components of the superconductors. In this way, time reversal symmetry breaking can be established and all parameters needed to fully describe the pair potential can be fully determined even in the absence of time reversal symmetry.

The paper is setup as follows. In Sec. \ref{sec:Model} we present the model for an SFN junction with time-reversal and inversion symmetry broken superconductors, in Sec. \ref{sec:Helical} we show the conductance calculated using this model for (i)s + helical p-wave superconductors. Next, in Sec. \ref{sec:chiral} we present the results for (i)s + chiral p-wave superconductors and compare them to those of (i)s + helical superconductors. We conclude our article in Sec. \ref{sec:Discussion} with a discussion of our results and discuss how to generalize to other types of non-centrosymmetric time-reversal broken superconductors.
\section{Model}\label{sec:Model}
\begin{figure}
    \centering
    \includegraphics[width = 8.6cm]{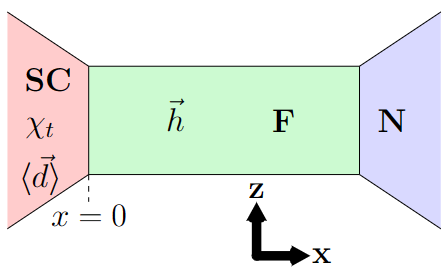}
    \caption{A schematic of the SFN junction. The superconductor (SC) is characterized by the phase difference $\chi_{t}$ between the singlet and triplet components and by the d-vector of the triplet component. The ferromagnet (F) is characterized by its exchange field. Both the SC and the normal metal (N) serve as electrodes.}
    \label{fig:Setup}
\end{figure}
We study a junction in which a ferromagnetic bar (F) is sandwiched between an unconventional superconductor and a normal metal electrode, schematically shown in Fig. \ref{fig:Setup}. Our model is similar to the one used in \cite{kokkeler2023proximity}, 
with the difference that an exchange interaction is included in the bar. We assume that the ferromagnetism is weak enough that we may use the quasiclassical formalism, $h/E_{F}\ll 1$. This assumption is valid  in weak ferromagnets or normal metals proximized by a ferromagnetic insulator. The induced exchange field causes Larmor precession of the electrons. Without proximity effect this does not affect the equilibrium properties of the metal, it only affects the distribution functions. However, if pair correlations exist, the presence of Larmor precession implies that while the spins of the pairs are still antiparallel when measured at the same time, they are in general not parallel when measured at different times. Thus, there is singlet-triplet conversion, where the d-vector of the triplets is parallel to the exchange field direction \cite{bergeret2018colloquium}. In S/F bilayers this leads to spin splitting of the superconductor \cite{bergeret2018colloquium}.  

 We assume that the scattering length in the problem is much smaller than any other relevant length, except the Fermi length, as is likely in thin films \cite{dingle1950electrical,gantmakher2005electrons}. In this case the Green's function is almost isotropic, and the Keldysh Usadel formalism \cite{belzig1999quasiclassical,chandrasekhar2008proximity} may be used to describe the F bar. We use the basis $(\psi_{\uparrow},\psi_{\downarrow},\psi^{\dagger}_{\downarrow},-\psi^{\dagger}_{\uparrow})^{T}$.
 If the width of the bar is either much larger or much smaller than the thermal coherence length, an effectively one-dimensional model may be used. In this limit the Green's function $G$ is approximately independent of the y and z coordinates:
 %Moreover, we assume that the bar is either very wide or very narrow in the directions perpendicular to the transport direction, so that an effectively one-dimensional Usadel equation may be used to describe the system:
\begin{align}
    D\partial_{x}(\bar{G}\partial_{x} \bar{G}) = [i(E+\vec{h}\cdot\vec{\sigma})\tau_{3},\bar{G}],
\end{align}
where $D$ is the spatially invariant diffusion constant, $\vec{\sigma}$ is the vector of Pauli matrices in spin space, $G$ is the isotropic component of the Green's function, $E$ is energy and $\vec{h}$ is the exchange field in the ferromagnet, with magnitude $h$.
If the interface resistance of the F/N interface is very small, for example if the F bar is a normal metal proximized by an FI and the N electrode is the same normal metal, the Green's function is continuous at the F/N interface:
\begin{align}
    \bar{G}(x = L) = \bar{G}_{N},
\end{align}
where $\bar{G}_{N} = \begin{bmatrix}
    \check{G}_{N}^{R}&\check{G}_{N}^{K}\\0&\check{G}_{N}^{A}
\end{bmatrix}$ is the Green's function in the normal metal electrode, where the retarded Green's function is $\check{G}_{N}^{R} = \tau_{3}$, the advanced Green's function is $\check{G}_{N}^{A} = -\tau_{3}$. The Keldysh Green's function is given by $\check{G}^{K} = \check{G}^{R}\check{h}-\check{h}\check{G}^{A}$, where $\check{h} = \hat{f}_{L0}\otimes\tau_{0}+\hat{f}_{T0}\otimes\tau_{3}$ is the matrix distribution function, with the longitudinal (L) and transverse(T) components \cite{schmid1975linearized,heikkila2019thermal}  determined by the Fermi-Dirac distribution: $\hat{f}_{L0,T0} = \frac{1}{2}\left(\tanh\frac{E+eV}{2k_{B}T}\pm\tanh\frac{E-eV}{2k_{B}T}\right)$, where $V$ is the voltage applied to the normal metal electrode, $T$ is the temperature of the system, which we assume to be well below $T_{c}$ and $k_{B}$ is the Boltzmann constant. 

%\TK{I remembered that the question about the inverse proximity effect was also asked in my paper with Sebastian Bergeret and Alberto Hijano, then we realized the assumption of weak contact/weak ferromagnetism is not even needed since we treat the S and N as electrodes from the start. The only assumption on the strength of the ferromagnet is that it is weak enough to use quasiclassics.} 
We assume that both the S and the N junctions are electrodes, while the F layer is a restriction between these two. In that case the inverse proximity effect of the ferromagnet on the (i)s + p-wave superconductor may be ignored, and the Tanaka-Nazarov boundary conditions \cite{tanaka2003circuit,tanaka2004theory,tanaka2022theory}, the extension of Nazarov's boundary conditions \cite{nazarov1999novel} to interfaces with unconventional superconductors, may be used. The boundary condition at the S/F interface reads:
\begin{equation}\label{eq:Tanaka-Nazarov}
    \bar{G}\nabla \bar{G}(x=0) = \frac{1}{\gamma_{BS} L}\langle \bar{S}(\phi)\rangle \; ,
\end{equation}
where
\begin{align}
    \bar{S}(\phi) &= \Tilde{T}(1+T_{1}^{2}+T_{1}(\bar{C}\bar{G}+\bar{G}\bar{C}))^{-1}(\bar{C}\bar{G}-\bar{G}\bar{C})\; ,\\
    \bar{C} &=\bar{H}_{+}^{-1}(\bar{\mathbf{1}}-\bar{H}_{-})\; ,\\
    \bar{H}_{+}&=\frac{1}{2}(\bar{G}_{S}(\phi)+\bar{G}_{S}(\pi-\phi))\label{eq:Hplus}\; ,\\
    \bar{H}_{-}&=\frac{1}{2}(\bar{G}_{S}(\phi)-\bar{G}_{S}(\pi-\phi))\label{eq:Hmin}\;.
\end{align}
Averaging over all modes passing through the interface is denoted by $\langle\cdot\rangle$, $\gamma_{BS} = R_{B}/R_{d}$ is the ratio between the boundary resistance and the normal states resistance of F and the parameter $T_{1} = \Tilde{T}/(2-\Tilde{T}+2\sqrt{1-\Tilde{T}})$ \cite{tanaka2004theory}, where $\Tilde{T}$ is the interface transparency given by 
\begin{align}\label{eq:zdef}
\Tilde{T}(\phi) = \frac{\cos^{2}\phi}{\cos^{2}{\phi}+z^{2}}\; ,
\end{align} 
with $z$ the Blonder-Tinkham-Klapwijk (BTK) parameter.
The retarded part of the Green's function of the superconductor $G_{S}(\phi)$ is given by the bulk equilibrium Green's function of an (i)s + p-wave superconductor.
For is + p-wave superconductors the pair potential is given by
\begin{align}
    \Delta(\phi) = e^{i\frac{\pi}{2}\chi_{t}} \frac{1}{\sqrt{r^{2}+1}}+\frac{r}{\sqrt{r^{2}+1}}\vec{d}(\phi)\cdot\vec{\sigma},
\end{align}
where $\vec{d}(\phi)$ is the d-vector an angle dependent unit vector that is different for different types of p-wave superconductors, $\chi_{t}$ is the phase difference between the singlet and triplet components, and $r$ is the mixing parameter. Therefore, the bulk Green's function reads
%$\bar{G}_{S}(\phi)$ denotes the equilibrium bulk Green's function of an is + p-wave superconductor, with retarded part
\begin{align}
    \check{G}^{R}_{S}(\phi) &= \frac{1}{2}(1+\vec{\hat{d}}(\phi)\cdot\vec{\sigma})\otimes\frac{1}{\sqrt{E^{2}-|\Delta_{+}|^{2}}}\begin{bmatrix}
        E&\Delta_{+}\\-\Delta_{+}^{*}&-E
    \end{bmatrix}\nonumber\\&+\frac{1}{2}(1-\vec{\hat{d}}(\phi)\cdot\vec{\sigma})\otimes\frac{1}{\sqrt{E^{2}-|\Delta_{-}|^{2}}}\begin{bmatrix}
        E&\Delta_{-}\\-\Delta_{-}^{*}&-E
    \end{bmatrix}\; ,\nonumber\\
    \Delta_{\pm} &= \frac{e^{i\frac{\pi}{2}\chi_{t}}\pm re^{i\psi(\phi)}}{\sqrt{r^{2}+1}}\;,
\end{align}
while its distribution function is the equilibrium Fermi-Dirac distribution. The advanced and Keldysh components can now be found using respectively  $\check{G}^{A}_{S} = -\tau_{3}\left(\check{G}^{R}_{S}\right)^{\dagger}\tau_{3}$ and $\check{G}_{S}^{K} = \check{G}^{R}_{S}\check{h}-\check{h}\check{G}^{A}_{S}$, with $\check{h} = \hat{f}_{L}\mathbf{1}+\hat{f}_{T}\tau_{3}$.
The following set of parameters is used throughout the paper: $\gamma_{BS} = 2$, $z = 0.75$, $E_{\text{Th}}/\Delta_{0} = 0.02$ to compare with the results on non-centrosymmetric superconductors in previous articles \cite{kokkeler2023proximity},\cite{kokkeler2023anisotropic}.

\section{Helical superconductors}\label{sec:Helical}
\begin{figure*}
    \centering
    {\hspace*{-1.5em}\includegraphics[width =8.6cm]{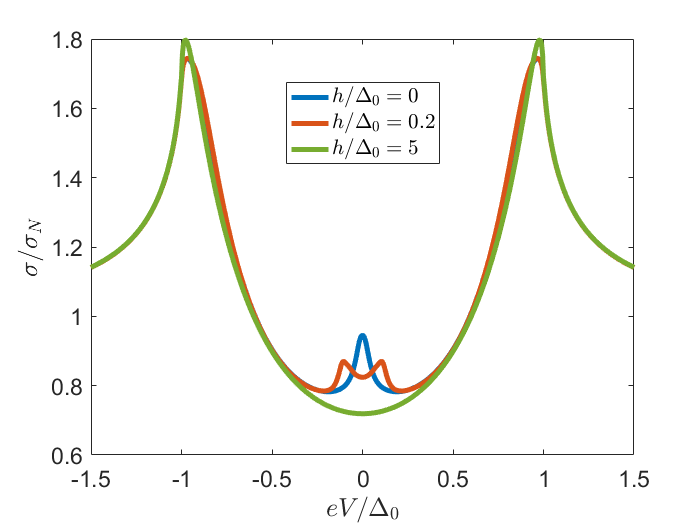}}
    \hfill
    {\hspace*{-2em}\includegraphics[width =8.6cm]{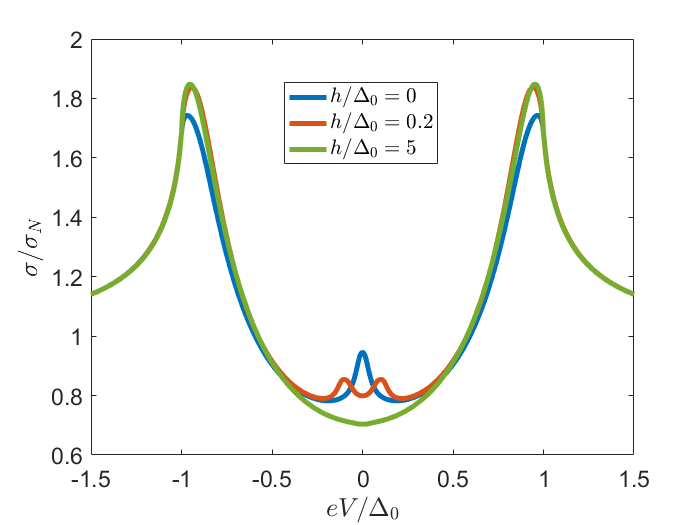}}
    \hfill
    {\hspace*{-1.5em}\includegraphics[width = 8.6cm]{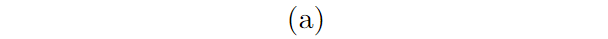}}
    \hfill
    {\hspace*{-1.5em}\includegraphics[width = 8.6cm]{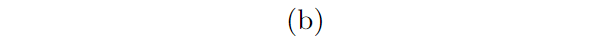}}
    \hfill
    \caption{The magnetic field dependence of the conductance for $\chi_{t} = 1$ for parallel (a) and perpendicular (b) orientations when $r = 0.5$. The sharp peak that without magnetic field is at zero bias splits into two peaks at $|eV| = h$ for small $h$, disappearing completely for large $h$. For voltages just below $\Delta_{0}$ the conductance is enhanced by the exchange field. This effect is stronger for the perpendicular orientation than for the parallel orientation.}\label{fig:hdep}
\end{figure*}
In the first part we focus on (i)s + helical p-wave 
superconductors, that is, the p-wave component has d-vector
\begin{align}
    \vec{d}(\phi) = (\cos{\phi},\sin{\phi},0).
\end{align}
We contrast the results for is + helical p-wave ($\chi_{t} = 1$) superconductors with those obtained for s + helical p-wave ($\chi_{t} = 0$) superconductors, and then provide an explanation for their differences.

\begin{figure*}
    \centering
    {\hspace*{-1.5em}\includegraphics[width =8.6cm]{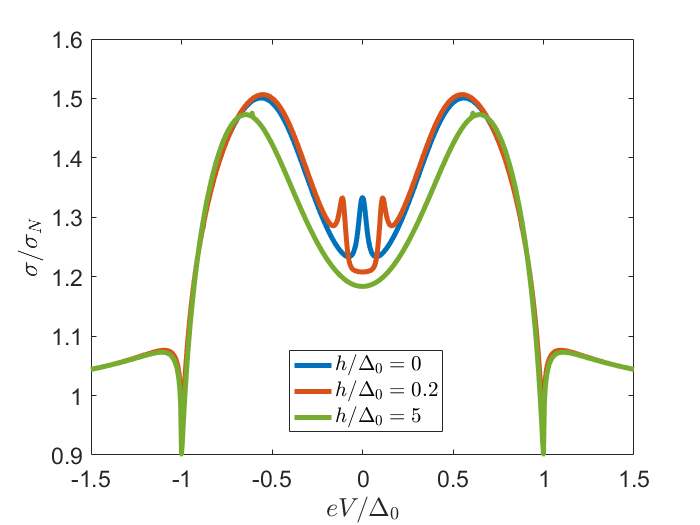}}
    \hfill
    {\hspace*{-2em}\includegraphics[width =8.6cm]{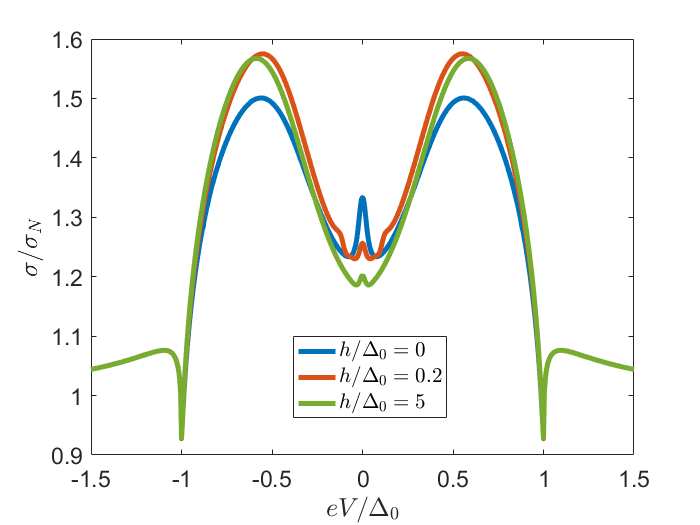}}
    \hfill
    \includegraphics[width = 8.6cm]{figures/A.png}
    \hfill
    \includegraphics[width = 8.6cm]{figures/B.png}
    \hfill
    \caption{The magnetic field dependence of the conductance for $\chi_{t} = 1$ and $r = 2$ for parallel (a) and perpendicular (b) orientations of the exchange field with respect to the d-vector. The sharp peak that without magnetic field is at zero bias splits into two peaks at $|eV| = h$ for small $h$, disappearing for large $h$. For voltages just below $\Delta_{0}$ the conductance is enhanced by the exchange field. This effect is stronger for the perpendicular orientation than for the parallel orientation.}\label{fig:hdepperp}
\end{figure*}

The exchange field dependence of the conductance of an SFN junction with is + helical p-wave superconductors is shown in Fig. \ref{fig:hdep} for s-wave dominant superconductors ($r = 0.5$) and Fig. \ref{fig:hdepperp} for p-wave dominant superconductors ($r = 2$). In both Figs. \ref{fig:hdep} and \ref{fig:hdepperp} the left panel corresponds to a parallel $\vec{h}$ and $\langle\vec{d}\rangle$, while the right panel corresponds to the perpendicular orientation of these two vectors. In all panels, there is a zero bias conductance peak with a width of the order of the Thouless energy for $h = 0$. This peak is due to coherent Andreev reflection and is split into two peaks at $|eV| = h$ for all orientations of the field, even if the p-wave component is dominant, in contrast to the s+p-wave superconductors. For s+p-wave pairing,
the topological property, i.e. the presence of ZES felt by the 

perpendicular injected  quasiparticles
does not change as far as  $\Delta_{p} > \Delta_{s}$ \cite{kokkeler2023anisotropic}.
On the other hand, for the is+p-wave superconductor, even if the magnitude of the s-wave
component is infinitesimal, ZBCP splits. This is due to the presence of both singlet and triplet components at $E = 0$ in the presence of time-reversal symmetry breaking \cite{kokkeler2023proximity}. In the p-wave dominant case a small zero energy peak remains if $\vec{h}$ and $\langle \vec{d}\rangle$ are perpendicular due to the long range triplets \cite{bergeret2001longrange}, see Fig. \ref{fig:hdepperp}(b), in all other cases the zero bias conductance peak is converted into a zero bias conductance dip, see Fig. \ref{fig:hdep} and \ref{fig:hdepperp}(a).
The height of the peak decreases for increasing exchange field strength, for $h\gg\Delta_{0}$ it is diminished.

If the s-wave component is dominant, $r<1$, shown in Fig. \ref{fig:hdep}, there is a sharp peak in the conductance around $|eV| = \Delta_{0}$. This peak, which is due to a suppression of the boundary resistance, is enhanced in the presence of an exchange field. If the field is parallel to the d-vector of the p-wave component this enhancement is only apparent for large fields, i.e. $h\gg\Delta_{0}$, as shown in Fig. \ref{fig:hdep}(a). On the other hand, if $\vec{h}$ is perpendicular to $\langle \vec{d}\rangle$, a small field is enough to increase the conductance of the junction, as shown in Fig. \ref{fig:hdep}(b). In both cases the conductance does not increase indefinitely, but saturates at a value that is slightly higher than for the perpendicular orientation than for the parallel orientation. The maximum increase of conductance is of the order of $0.1\sigma_{N}$.

If the p-wave superconductor is dominant, $r>1$, the peak below $|eV| = \Delta_{0}$ is much wider, as shown in Fig. \ref{fig:hdepperp}. Also in this case the conductance is enhanced by an exchange field. In the parallel orientation this enhancement only appears in a small window close to $|eV| = \Delta_{0}$ and only for large fields, see Fig. \ref{fig:hdepperp}(a). For the perpendicular orientation on the other hand the enhancement appears in a large window, and is largely enhanced compared to parallel field, for small fields $h\ll\Delta_{0}$. If the field is parallel a conductance enhancement of a few percent can be achieved, for the perpendicular orientation an enhancement of $0.1\sigma_{N}$ can be achieved, as in the case $r<1$.
%There is a notable increase of the conductance with exchange field for $eV$ just below $\Delta$r. Such increase is known to be possible in junctions in which ferromagnets are present \cite{Ozaeta2012andreev}, but does not appear in junctions with s + p-wave superconductors \cite{kokkeler2023proximity}.
%The enhancement of conductance is weak if $\vec{h}$ and $\langle \vec{d}\rangle$ are parallel, Fig. \ref{fig:hdep}, but strong if they are perpendicular, Fig. \ref{fig:hdepperp}.
%This implies that there is a strong dependence of the conductance on the direction of the field. This is quite different from the magnetic field dependence in s + p-wave junctions.
Comparing this to the magneto-electric response in s + helical p-wave superconductors \cite{kokkeler2023anisotropic}, we find that the magneto-electric response in is + helical p-wave superconductors has opposite sign and a much stronger anisotropy.
The difference between these materials is due to the presence or absence of time-reversal symmetry. For s + p-wave superconductors, time-reversal symmetry is not broken. Moreover, in a dirty material, the scattering rate is high and therefore the Green's function is almost isotropic. Therefore, the density of states in the surface Green's function $C$ is independent of spin. On the other hand, for the is + p-wave superconductor, time-reversal symmetry is broken, and the surface density of states is different for spins parallel or antiparallel to $\langle\vec{d}\rangle$. 

%Therefore, in the presence of time-reversal symmetry the density of states in the dirty normal metal is required to be spin independent. On the other hand, for the is + p-wave superconductor, time-reversal symmetry is broken and the density of states is different for different spin-orientations. In fact, the is + p-wave superconductor the density of states is similar to that of a spin-split superconductor.

Mathematically, based on this the conductance enhancement can be inferred from the Tanaka-Nazarov boundary conditions, specifically, the Keldysh component \cite{tanaka2004anomalous}. If the density of states of the surface Green's function has no spin-dependence, as for the s+p-wave superconductor, the only difference in the commutator or anticommutator caused by the exchange field comes from the terms off-diagonal in Nambu space. Since the pair amplitude in the normal metal is suppressed by a finite boundary resistance, the off-diagonal terms are small and hence the effect  of the exchange field on the anticommutator is small in this case.

On the other hand, for the is + p-wave case  also the density of states of the surface Green's function has a spin dependence and also the contribution of the density states to the anticommutator in the Tanaka-Nazarov boundary condition depends on the exchange field. This greatly reduces the anti commutator $CG+GC$. This implies that the enhancement of the conductance is largest for those voltages for which $T_{1}(CG+GC)$ is the dominant term in the denominator. As shown in  \cite{kokkeler2023proximity}, for is + p-wave superconductors this term is dominant for $\Delta_{s}<|eV|<\sqrt{\Delta_{s}^{2}+\Delta_{p}^{2}}$, where we define $\Delta_{s,p}$ as the magnitudes of the singlet and triplet pair amplitudes. Thus, the enhancement the conductance is most prominent in a broader voltage window if the p-wave component of the pair potential larger. 

To understand the physical mechanism behind the enhancement of conductance just below $|eV| = \Delta_{0}$ we first consider the physical interpretation of the denominator. From the derivation in \cite{nazarov1999novel} it follows that the denominator can be attributed to higher order tunnelling. Indeed, to first order in the tunnelling $T_{1}$ parameter between two materials with Green's functions $G_{1,2}$ the current between these materials is given by $T_{1}[G_{1},G_{2}]$, the well-known Kuprianov-Luckichev boundary condition \cite{kuprianov1988influence}. The boundary condition involves a commutator, which is larger if the Green's functions of the two materials are more different, for example in an SIS junction if the phase difference between two superconductors is larger.

It is only upon inclusion of the higher orders in $T_{1}$ that the denominator appears, which suppresses the first order approximation.  The Green's function exactly at the interface is altered by the tunnelling of the electrons of the other material into it. Therefore the difference in Green's functions on either side of the interface is smaller, leading to a reduction of the current compared to the first order estimation. 

%Based on this we may contrast the case in which the exchange field is perpendicular to the d-vector with the case in which they are parallel. 
Next we consider the effect of the relative orientation of the d-vector in the two materials on these higher order terms. Suppose a fraction $\alpha$ of electrons is exchanged between materials 1 and 2 due to the tunnelling. If the spin quantization axis is the same for both materials 1 and 2, the resulting new pair amplitudes and density of states are a weighted average of the two Green's functions. 

This is not the case if the spin quantization axes are perpendicular, for example along the z-axis in material 1 and along the x-axis in material 2. The electrons moving from 2 to 1 do contribute to all spin-averaged properties, i.e. to those parts of the Green's function proportional to $\sigma_{0}$.   However, the electrons that move from 2 to 1 do not have any spin-projection along the z-axis and therefore can not contribute to the $\sigma_{z}$-terms in the Green's function of material 1. %These terms are altered because a fraction $\alpha$ of the electrons in material 1 has been removed from there. %There is a reduction of these terms since after the exchange only a fraction $1-\alpha$ of the electrons contributes to it. 
Thus, the change in the $\sigma_{z}$-term of the Green's function in 1 is smaller in the perpendicular orientations. Likewise, the electrons moving from material 1 to 2 can not contribute to the $\sigma_{x}$-terms in the Green's function of material 2, because the electrons from material 1 do not have any spin-projection along the x-direction.
Therefore, if the spin quantization axis is different in the two materials, the Green's functions at the interface are more different compared to the case in which they are parallel. For this reason the current is minimized if the d-vectors are parallel.

\begin{figure*}
    \centering
    {\hspace*{-1.5em}\includegraphics[width = 8.6cm]{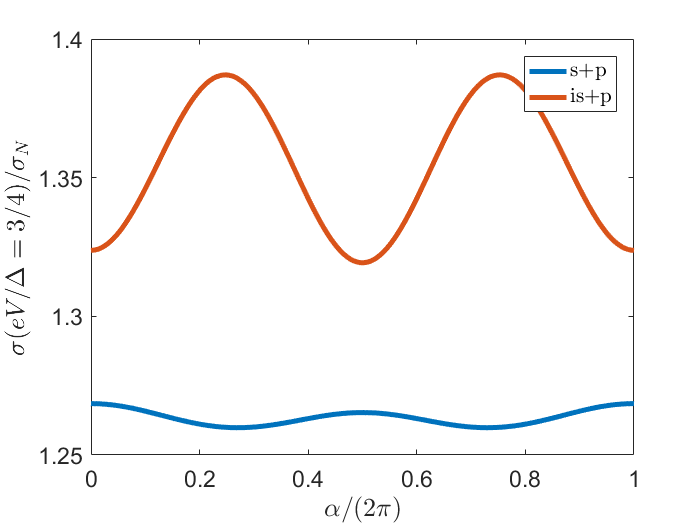}}
    \hfill
    {\hspace*{-2em}\includegraphics[width = 8.6cm]{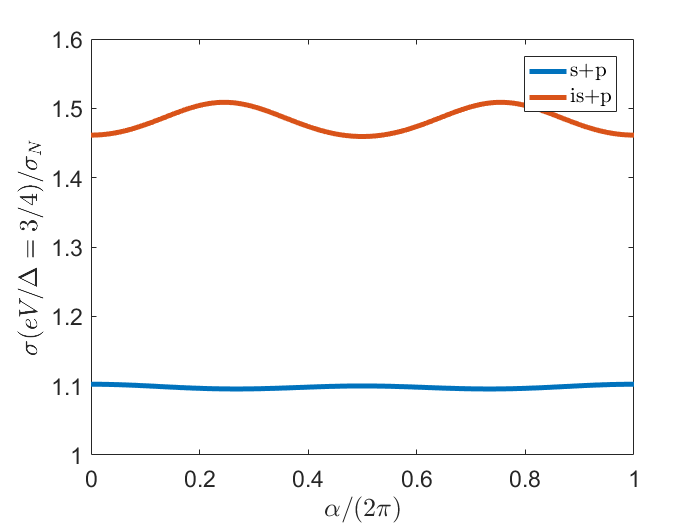}}
    \hfill
    \includegraphics[width = 8.6cm]{figures/A.png}
    \hfill
    \includegraphics[width = 8.6cm]{figures/B.png}
    \hfill
    \caption{The quadrupolar dependence of the differential conductance at $eV/\Delta_{0} = 0.75$ on the angle between the exchange field and d-vector of the superconductor for $r = 0.5$ and $r = 2$. The exchange field strength is $h/\Delta_{0} = 0.2$. For s + helical p-wave superconductors the conductance is maximized for the parallel orientation. On the other hand, for is + helical p-wave superconductors the conductance is minimized in the parallel orientation and the angular dependence is much larger. There is also a small bipolar component, $\sigma(\alpha = \pi)\neq \sigma(\alpha = 0)$, however the quadrupolar component dominates.}
    \label{fig:alphadeps}
\end{figure*}

In the setup used in this paper, % if the d-vector and exchange field are parallel, the d-vectors in the superconductor and the bar are necessarily also parallel. On the other hand, 
if the exchange field is not parallel to the d-vector it rotates the quantization axes of the induced pair correlations in the ferromagnet. Meanwhile, since the exchange field is only present in the normal metal the quantization axis of the superconductor is unchanged. Therefore, the current increases if the exchange field is not parallel to the d-vector of the superconductor, explaining the enhancement of the conductance compared to the normal state and the anisotropy in the enhancement. This leads to a quadrupolar dependence of the conductance on the exchange field, as shown in Fig. \ref{fig:alphadeps}. A quadrupolar dependence on the exchange field in proximity structures has been found before, both in supercurrents \cite{bergeret2001enhancement} and conductance \cite{bergeret2001longrange}. Our effect however differs from \cite{bergeret2001longrange} via the voltage range in which the effect appears.

 Though smaller than for the perpendicular orientation, for the s-wave dominant case there is also an enhancement of conductance just below $|eV| = \Delta_{0}$ in the parallel orientation compared to the case with $h = 0$ as shown in Fig. \ref{fig:hdep}(a). Even in the p-wave dominant case there is a very small enhancement in this voltage window, see Fig. \ref{fig:hdepperp}(a). This is a specific feature of the helical p-wave superconductor. Indeed, for the helical p-wave superconductor the direction of the d-vector is parallel to momentum, and therefore even if $\vec{h}$ and $\langle\vec{d}\rangle$ are parallel there exist modes for which $\vec{h}$ and $\vec{d}(\phi)$ are not, and the conductance is enhanced. Since the mode at normal incidence has the largest transmission eigenvalue, 
 %the enhancement of conductance due to modes with oblique incidence is small. Therefore, 
 the enhancement in conductance if $\vec{h}\parallel\langle\vec{d}\rangle$ is small compared to the enhancement for $\vec{h}\perp\langle\vec{d}\rangle$. 
%This indeed leads to the presence of an anticommutator upon normalization.

The predicted sign and anisotropy of the magnetoresistance can only be found if the exchange field strength $h$ is much smaller than the Fermi energy $E_{F}$.  If the ferromagnetic interaction is significant compared to the Fermi energy, the difference in momenta for opposite spins is large. In that case the conductance is maximized for the parallel orientation and minimized for the antiparallel orientation in F/S/F or F/F junctions \cite{bozovic2002coheren}\cite{miyazaki1995giant}. In our formalism such effects are suppressed by a factor $h/E_{F}$ and therefore negligible compared to the effect described here as long as $h/E_{F}\ll 1$. The difference in symmetries between the two effects 
in fact allows one to disentangle them. Also within our quasiclassical formalism there is a small bipolar component, however this component is small compared to the quadrupolar component.
%Now consider two superconductors in which the spin-polarization in the density of states is in the same direction. In each of the superconductors a different fraction of electrons is condensed into Cooper pairs. If a fraction $T$ of the electrons in material 2 moves into material 1, the resulting fraction of electrons cond
%Now we may make an analogy with a junction between two strong ferromagnets. First consider the case in which they are antiparallel. If  

%If we add two parallel vectors, one of which a factor $T$ smaller than the other, the length of those vectors is much $1\pm T$. On the other hand, if the two vectors are perpendicular, the resulting length is $\sqrt{1+T^{2}}$, that is, 

\begin{figure*}
    \centering
    {\hspace*{-1.5em}\includegraphics[width =8.6cm]{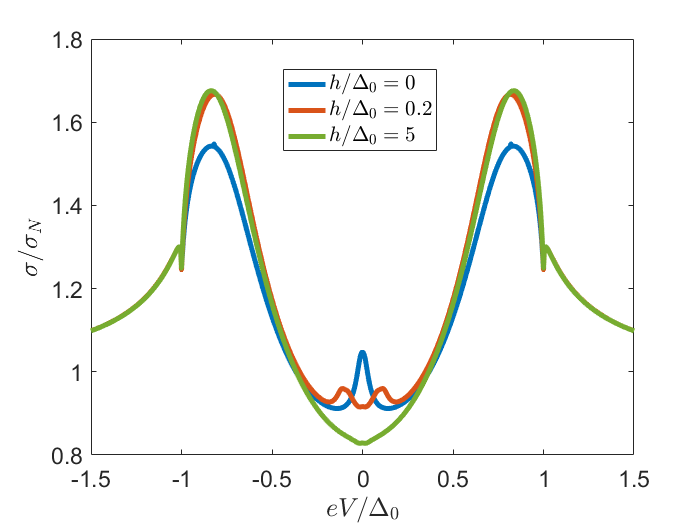}}
    \hfill
    {\hspace*{-2em}\includegraphics[width =8.6cm]{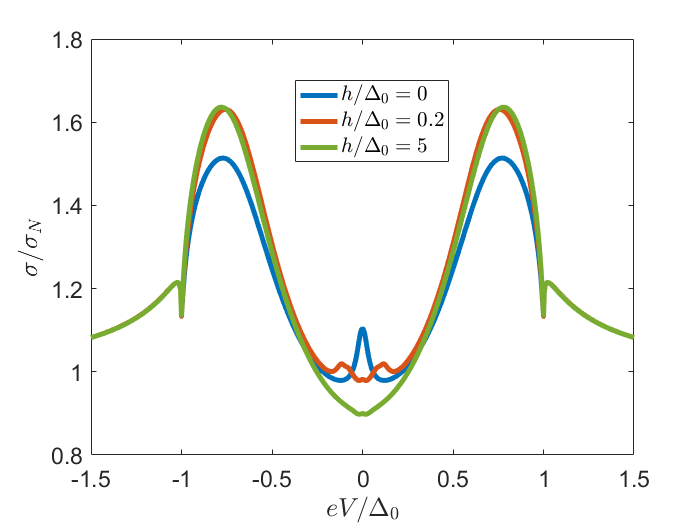}}
    \hfill
    \includegraphics[width = 8.6cm]{figures/A.png}
    \hfill
    \includegraphics[width = 8.6cm]{figures/B.png}
    \hfill
    \caption{The magnetic field dependence of the conductance for $\chi_{t} = 1$ for $r = 0.9$ (left) and $r = 1.1$ (right) when the field is perpendicular to the d-vector. The sharp peak that without magnetic field is at zero bias splits into two peaks at $|eV| = h$ for small $h$, dissappearing completely for large $h$. The increase in the conductance by applying a magnetic field is larger than for $r = 0.5$ or $r = 2$, confirming that it is an effect for which the coexistence of even and odd-parity contributions to the pair potential.}\label{fig:hdepperp3}
\end{figure*}

Because time-reversal symmetry is broken in the superconductor, for is + helical p-wave superconductors $r = 1$ is not a topological phase transition, unlike the s + helical p-wave superconductor \cite{sato2017topological,schnyder2008classification,leijnse2012introduction,tanaka2009theory}. Therefore the conductance is continuous as a function of $r$, both in the absence and in the presence of an exchange field, as confirmed by the results in Fig \ref{fig:hdepperp3}.

\section{Chiral superconductors}\label{sec:chiral}
\begin{figure*}
    \centering
    {\hspace*{-1.5em}\includegraphics[width = 8.6cm]{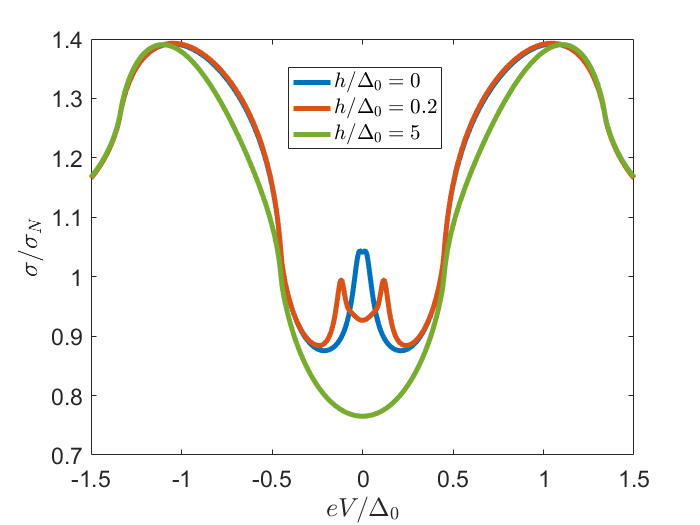}}
    \hfill
    {\hspace*{-2em}\includegraphics[width = 8.6cm]{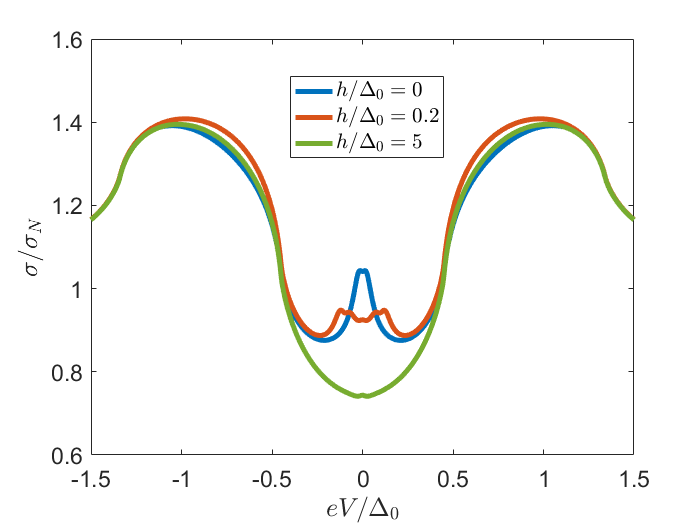}}
    \hfill
    \includegraphics[width = 8.6cm]{figures/A.png}
    \hfill
    \includegraphics[width = 8.6cm]{figures/B.png}
    \caption{The conductance in is + chiral p-wave junctions in the presence of parallel (left) and perpendicular (right) exchange fields. The presence of a small exchange field can increase the conductance in the junction. The increase is less visible then for is + helical p-wave junctions because the phase is mode-dependent. Therefore, only the conductance of a selection of modes is enhanced, while that of the others is suppressed. The superconductor is s-wave dominant, with $\Delta_{p}/\Delta_{s} = 0.5$.}\label{fig:R0p5TRB1}
\end{figure*}

\begin{figure*}
    \centering
    {\hspace*{-1.5em}\includegraphics[width = 8.6cm]{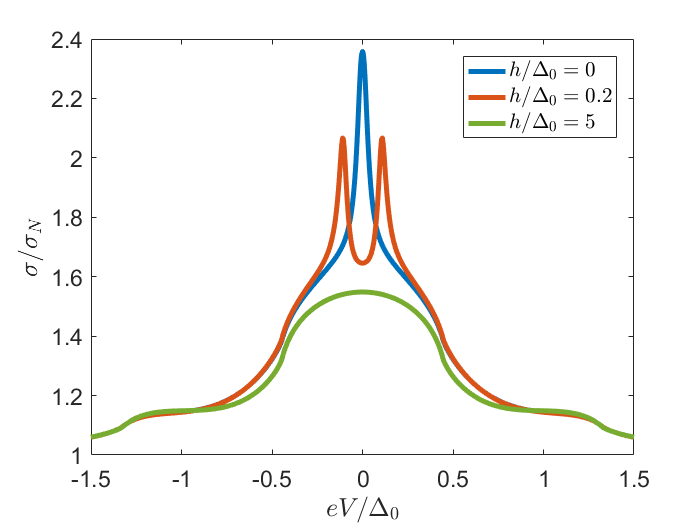}}
    \hfill
    {\hspace*{-2em}\includegraphics[width = 8.6cm]{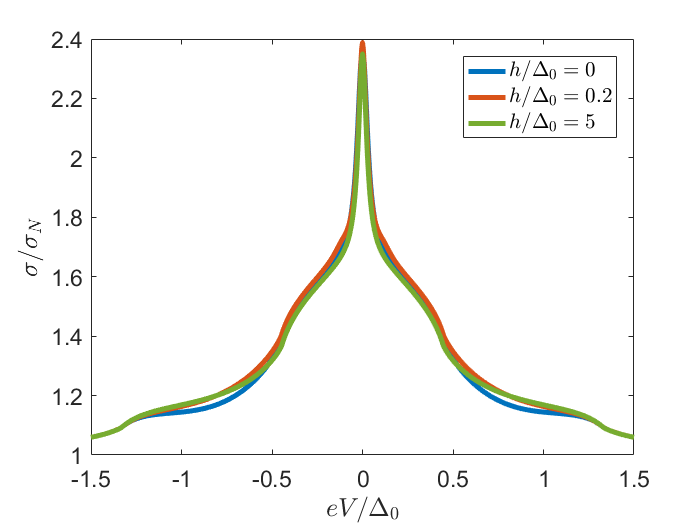}}
    \hfill
    \includegraphics[width = 8.6cm]{figures/A.png}
    \hfill
    \includegraphics[width = 8.6cm]{figures/B.png}
    \caption{The conductance in is + chiral p-wave junctions in the presence of parallel (left) and perpendicular (right) exchange fields. The superconductor is p-wave dominant, with $\Delta_{p}/\Delta_{s} = 2$. The sharp zero bias conductance peak, splits upon application of a parallel exchange field, but hardly changes by application of a perpendicular field. This happens because peak is due to the oblique modes, for which the phase difference between the modes is almost 0.}\label{fig:R2TRB1}
\end{figure*}

The same calculations were performed for chiral superconductors, for which the d-vector is given by
\begin{align}
    \vec{d}(\phi) = e^{i\phi}(0,0,1).
\end{align}
We used the same set of parameters as for the junction with (i)s + helical p-wave superconductors, that is, $\gamma_{BS} = 2$, $z = 0.75$, $E_{\text{Th}}/\Delta_{0} = 0.02$. The dependence of the conductance on the exchange field is illustrated in figure \ref{fig:R0p5TRB1} for s-wave dominant superconductors and \ref{fig:R2TRB1} for p-wave dominant superconductors. We first elaborate on the results and how they depend on parameters, then we provide a physical explanation for the differences obtained with the (i)s + helical p-wave junctions.

%If the s-wave component is dominant, the zero bias conductance is suppressed by an exchange field both if it is parallel and perpendicular to the d-vector. For small exchange fields, the zero bias peak splits into two peaks at $|eV|\approx h$, as for the helical superconductors.
%For larger voltages, the conductance is suppressed by an exchange field for s + p-wave superconductors, regardless of $r$, as shown in Fig. \ref{fig:R0p5TRB0}. This conductance suppression is of the order of 5\% of the normal state conductance for the perpendicular orientation, see Fig. \ref{fig:R0p5TRB0}(b), for the parallel orientation it is almost negligible for small fields and at most of the order of 1\% for $h\gg\Delta_{0}$ (Fig. \ref{fig:R0p5TRB0}(a)).  
Also for is + chiral superconductors a weak exchange field leads to an increase in the conductance for all $r$, see Fig. \ref{fig:R0p5TRB1}. As for the helical case, the enhancement is much stronger if the exchange field is perpendicular to the d-vector (Fig. \ref{fig:R0p5TRB1}(b)) than if it is parallel to it (Fig. \ref{fig:R0p5TRB1}(a)). The enhancement however, is smaller than for the s + helical p-wave superconductors, maximally around $0.05\sigma_{N}$ for the perpendicular orientation and almost negligible in the parallel orientation, and is suppressed for larger exchange fields, as shown in \ref{fig:R0p5TRB1}(b). 
\begin{comment}
\begin{figure*}
    \centering
    {\hspace*{-1.5em}\includegraphics[width = 8.6cm]{figures/R0p5hang0TRB0Double.png}}
    \hfill
    {\hspace*{-2em}\includegraphics[width = 8.6cm]{figures/R0p5hang100TRB0Double.png}}
    \hfill
    \includegraphics[width = 8.6cm]{figures/A.png}
    \hfill
    \includegraphics[width = 8.6cm]{figures/B.png}
    \caption{The conductance in s + chiral p-wave junctions in the presence of parallel (left) and perpendicular (right) exchange fields. The conductance around $|eV| = \Delta_{0}$ is suppressed by application of an exchange field. The superconductor is s-wave dominant, with $\Delta_{p}/\Delta_{s} = 0.5$.}\label{fig:R0p5TRB0}
\end{figure*}
\end{comment}
\begin{comment}
\begin{figure*}
    \centering
    {\hspace*{-1.5em}\includegraphics[width = 8.6cm]{figures/R2hang0TRB0Double.png}}
    \hfill
    {\hspace*{-2em}\includegraphics[width = 8.6cm]{figures/R2hang100TRB0Double.png}}
    \hfill
    \includegraphics[width = 8.6cm]{figures/A.png}
    \hfill
    \includegraphics[width = 8.6cm]{figures/B.png}
    \caption{The conductance in s + chiral p-wave junctions in the presence of parallel (left) and perpendicular (right) exchange fields. The superconductor is p-wave dominant, with $\Delta_{p}/\Delta_{s} = 2$.}\label{fig:R2TRB0}
\end{figure*}
\end{comment}
%As shown in Fig. \ref{fig:R2TRB0}, for p-wave dominant s + chiral p-wave superconductors the zero bias conductance is only suppressed by a parallel exchange field (a), not by a perpendicular exchange field (b). 
For p-wave dominant is + chiral p-wave superconductors there is a weak suppression of the conductance in the presence of an exchange field (Fig. \ref{fig:R2TRB1}(b)), but it is very small compared to a parallel orientations, see Fig. \ref{fig:R2TRB1}(a). Correspondingly, the peaks at $|eV| = h$ only appear for parallel exchange fields. For $|eV|\approx\Delta_{0}$ the presence of an exchange field always decreases the conductance in junctions with s + p-wave superconductors, whereas it increases the conductance for is + p-wave superconductors, though the effect is again considerably smaller than for helical superconductors. Thus, while our method works well for s + helical p-wave superconductors, it does not do so if chiral superconductors are present.
\begin{figure}
    \centering
    \includegraphics[width = 8.6cm]{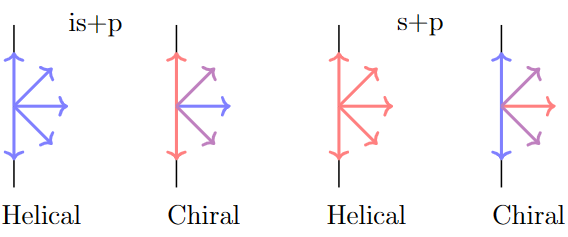}
    \caption{Schematic showing the difference between helical and chiral (i)s + p-wave superconductors. The phase difference between singlet and triplet superconductors $\text{arg}(\frac{\Delta_{p}(\phi)}{\Delta_{s}(\phi)})$ is illustrated by colour. For helical superconductors, this phase difference is independent of the direction of momentum, and hence all modes contribute to an enhancement of the conductance. On the other hand, for the chiral superconductor, the phase difference between the triplet and singlet components depends on the angle of momentum, and hence for some modes the conductance is enhanced while for others it is suppressed. }
    \label{fig:CartoonChiralHelical}
\end{figure}

The difference between the (i)s + chiral and (i)s + helical p-wave superconductors can be understood by considering the phase difference for each mode, depicted in Fig. \ref{fig:CartoonChiralHelical}. For (i)s + helical superconductors the phase difference between the s-wave and p-wave components is 0 or $\pi/2$ for all modes. On the other hand, in the chiral case the phase difference is mode dependent and therefore for some modes the conductance is enhanced, while for others it is suppressed. For s + chiral p-wave superconductors the phase difference ranges between $-\pi/2$ and $\pi/2$, whereas for is + p-wave superconductors it ranges between $-\pi$ and $0$. These two cases are different, because the transparency is higher for the mode at normal incidence compared to the modes at large angle incidences. However, the varying phase difference does considerably soften the effects, explaining why the observed enhancement is much weaker in (i)s + chiral superconductors than for (i)s + helical superconductors.

On the other hand, the anisotropy in the magnetoresistance is relatively much stronger for (i)s + chiral superconductors than for (i)s + helical superconductors. Indeed, for the chiral superconductor the direction of the d-vector is independent of the direction of momenta, and therefore for all modes exchange field and d-vector are either parallel or perpendicular. However, due to the much smaller magnetoresistance the absolute anisotropy is smaller than for the (i)s + helical superconductors.

The significant suppression of conductance for the is + chiral superconductors in the voltage region $\frac{1-r}{\sqrt{1+r^{2}}}<|eV|/\Delta_{0}<\frac{1}{\sqrt{1+r^{2}}}$ in Fig. \ref{fig:R2TRB1}(a) can also be understood using this picture. As discussed in the section on (i)s+helical p-wave superconductors, the modes for which the phase difference between singlet and triplet components is $\pi/2$ enhance the conductance only for $\Delta_{s}<|eV|<\sqrt{\Delta_{s}^{2}+\Delta_{p}^{2}}$, that is, $\frac{1}{\sqrt{1+r^{2}}}<|eV|/\Delta_{0}<1$. On the other hand, for modes in which the phase difference between singlet and triplet components is almost $0$, the conductance is suppressed in the larger window $\frac{1-r}{\sqrt{1+r^{2}}}<|eV|/\Delta_{0}<\frac{1+r}{\sqrt{1+r^{2}}}$.
Thus, for $\frac{1-r}{\sqrt{1+r^{2}}}<|eV|<\frac{1}{\sqrt{1+r^{2}}}$ the conductance is suppressed by an exchange field.

%while for s + helical p-wave case the conductance is enhanced above the normal state conductance for $\Delta_{s}-\Delta_{p}<eV<\Delta_{s}+\Delta_{p}$, for is + helical p-wave superconductors the enhancement is only for $\Delta_{s}<eV<\sqrt{\Delta_{s}^{2}+\Delta_{p}^{2}}$. Thus, in chiral superconductors, any enhancement for $\Delta_{s}-\Delta_{p}<eV<\Delta_{s}$ is due to modes for which the phase difference between the different components is 0. As argued before, for those modes the conductance is suppressed by application of an exchange field.

\begin{table*}[]
    \centering
    \begin{tabular}{|c|c|c|c|c|c|}
    \hline
       Type  & $\sigma(eV= 0)>\sigma_{N}$&$eV_{\text{ani,min}}/\Delta_{0}$&$eV_{\text{ani,max}}/\Delta_{0}$&$\sigma(h\neq 0)>\sigma(h =0)$&Minimization of conductance\\\hline
         s-wave&No&-&-&No&-\\\hline
         p-wave&Yes&-&-&No&-\\\hline
         s + p-wave, $r<1$&No&$\frac{1-r}{\sqrt{1+r^{2}}}$&$\frac{1+r}{\sqrt{1+r^{2}}}$&No&$\vec{h}\parallel\langle\vec{d}\rangle$\\\hline
         s + p-wave, $r>1$&Yes&$\frac{r-1}{\sqrt{1+r^{2}}}$&$\frac{r+1}{\sqrt{1+r^{2}}}$&No&$\vec{h}\parallel\langle\vec{d}\rangle$\\\hline
         is + p-wave, $r<1$&No&$\frac{1}{\sqrt{1+r^{2}}}$&1&Yes&$\vec{h}\perp\langle\vec{d}\rangle$\\\hline
         is + p-wave, $r>1$&Yes&$\frac{1}{\sqrt{1+r^{2}}}$&1&Yes&$\vec{h}\perp\langle\vec{d}\rangle$\\\hline
    \end{tabular}
    \caption{Signatures of the conductance in SNN/SFN junctions using (i)s + helical p-wave superconductors, based on the ratio of the zero bias conductance $\sigma(eV = 0)$ and normal state resistance $\sigma_{N}$ and the exchange field anisotropy in the voltage window $0<eV_{\text{ani,min}}<eV<eV_{\text{ani,max}}$. Each of the six categories has a unique combination of results in columns 2 to 5, with which the existence of mixed parity superconductivity, the dominant pair potential and time-reversal symmetry breaking may be detected. The parameter $r$ and the overall energy scale $\Delta_{0}$ can then be extracted using the results in columns 3 and 4 via $\text{min}(r,\frac{1}{r}) = \frac{eV_{\text{ani,max}}-eV_{\text{ani,min}}}{eV_{\text{ani,max}}+eV_{\text{ani,min}}}$ and $\Delta_{0}= \frac{eV_{\text{ani,max}}\sqrt{1+r^{2}}}{1+r}$, while the direction of $\langle\vec{d}\rangle$ may be extracted from column 6.}
    \label{tab:Identification}
\end{table*}
\section{Discussion}\label{sec:Discussion}
We have presented a method to determine the pair potential in time-reversal symmetry broken non-centrosymmetric superconductors. The differential conductance close to the gap edge is enhanced by an exchange field in the presence of time reversal symmetry breaking in the superconductor. Next to this, our results show that by varying the direction of the exchange field this enhancement can also be used to determine the direction of the d-vector of the triplet correlations. With this we have presented a theory for the complete determination of the pair potential in (i)s + helical p-wave superconductors, as summarized in Table \ref{tab:Identification}. The mixing parameter, the phase difference between singlet and triplet components and direction of the d-vector can all be determined from  conductance measurements in SFN junctions.

Our results show that if the exchange field is small compared to the Fermi energy, the magnetoresistance of the SFN junction is quadrupolar, and hence qualitatively different from the dipolar giant magnetoresistance found if the ferromagnetic interaction is comparable to the Fermi energy. Indeed, we found that maximal enhancement is obtained in perpendicular exchange fields, while minima are achieved in both parallel and antiparallel orientations, while GMR distinguishes between parallel and antiparallel orientations. Our effect is only present in a specific voltage window, determined by the mixing parameter, but can reach $0.1\sigma_{N}$ of the normal state conductance.

To determine the pair potential in an arbitrary time-reversal and inversion symmetry broken superconductor, one needs to include also the higher order angular momenta, d-wave f-wave etc. For such superconductors the conductance is different, but since the conductance enhancement depends only on the phase between the superconductors and the relative orientation of the spin in the superconductor and ferromagnet the qualitative features of our results should also be visible in s + f or d + p wave superconductors, as long as the individual components do not break time reversal symmetry, i.e. they are not chiral. For (i)s + chiral superconductors, which may be distinguished from (i)s + helical superconductors using the directional dependence of the d-vector \cite{kokkeler2023anisotropic} the mode dependence of the phase difference smoothens the results and makes the extraction of the parameter $r$ more difficult. However, $\chi_{t}$ may be determined using the conductance enhancement in the case $\chi_{t} = 1$. The theory can not be used for s +id (p+if) superconductors, as in such superconductors only singlet (triplet) correlations are present.
\section{Acknowledgements}
We would like to thank Sebastian Bergeret for usefull discussions.
T.K. acknowledges  financial support from Spanish MCIN/AEI/10.13039/501100011033 through project PID2020-114252GB-I00 (SPIRIT) and  TED2021-130292B-C42,
and the Basque Government through grant IT-1591-22. Y.T. acknowledges support from JSPS with
Grants-in- Aid for Scientific Research
( KAKENHI Grants No. 20H00131 and No. 23K17668).
\bibliography{biblio}
\end{document}